\documentclass[conference]{IEEEtran}
\usepackage{etoolbox}
\patchcmd{\thebibliography}{\section*{\refname}}{\section*{\refname}\vspace{-4pt}}{}{}
\renewcommand{\baselinestretch}{0.95}  
\usepackage{etoolbox}
\patchcmd{\thebibliography}{\section*{\refname}}{\section*{\refname}\vspace{-4pt}}{}{}
\renewcommand{\baselinestretch}{0.95}  
\usepackage{cite}
\usepackage{amsmath,amssymb,amsfonts}
\usepackage{algorithmic}
\usepackage{placeins} 
\usepackage[hidelinks]{hyperref}
\usepackage{url} 
\usepackage{graphicx}
\usepackage[caption=false]{subfig} 
\usepackage{array}
\usepackage{textcomp}
\usepackage{xcolor}

\makeatletter

\def\@SetAffiliation{}
\newcommand{\SetAffiliation}[1]{\gdef\@SetAffiliation{\small#1}}

\def\@SetAuthors{}
\newcommand{\SetAuthors}[1]{\gdef\@SetAuthors{#1}}

\def\@maketitle{%
	\newpage
	\null
	\vskip 2em%
	\begin{center}%
		\let \footnote \thanks
		{\bfseries\LARGE\sffamily \@title \par}%
		\vskip 1.5em%
		{
			\lineskip .5em%
			\begin{tabular}[t]{c}%
				{\@SetAuthors}
			\end{tabular}\par}
		\vskip 0.5em%
	\@SetAffiliation
		\vskip 1em%
		{\large \@date}%
	\end{center}%
	\par
	\vskip 1.5em}

\makeatother
\renewcommand\IEEEkeywordsname{Keywords-component} 
\usepackage{footnote}
\renewcommand{\figurename}{Figure.}
\renewcommand{\tablename}{TABLE.}
\begin{document}

\title{Bayesian Optimization of the GEKO Turbulence Model for Predicting Flow Separation Over a Smooth Surface}

\SetAuthors{
	Nikhila Kalia${^1*}$,
	Ryley McConkey$^2$,
	Eugene Yee$^1$,
	Fue-Sang Lien$^1$
	}

\SetAffiliation{
	$^1$Mechanical Engineering, University of Waterloo, 200 University Ave W, Waterloo, ON, Canada\\
	$^2$Research Laboratory of Electronics, Massachusetts Institute of Technology, 77 Massachusetts Avenue Cambridge, MA\\
    $*$Corresponding author: nkalia@uwaterloo.ca
	}

\date{February 15 2025}

\maketitle

\begin{abstract}
\normalfont
This paper extends the \textbf{tu}ning pa\textbf{r}ameters using \textbf{B}ayesian-\textbf{o}ptimization\textbf{-RANS} (turbo-RANS) methodology to enhance the predictive accuracy of Reynolds-averaged Navier-Stokes (RANS) turbulence models for flow over a converging-diverging channel, a benchmark case characterized by adverse pressure gradients and flow separation. Using Bayesian optimization, the generalized $k$-$\omega$ (GEKO) turbulence model was calibrated by tuning key coefficients ($C_\text{SEP}$ and $C_\text{NW}$) with sparse reference data from direct numerical simulation (DNS) studies at $Re = 12,600$. The optimized model was then tested on the streamwise velocity predictions against both DNS ($Re = 12,600$) and large-eddy simulation (LES) data, along with the coefficient of friction ($C_f$) predictions from the LES dataset at $Re = 20,580$. Results with optimized coefficient values of $C_\text{SEP}$ = 0.489 and $C_\text{NW}$ = 1.778 demonstrate significant improvements in the prediction of wall quantities, addressing key limitations of default RANS coefficients. These findings provide further insight into the application of machine learning-assisted RANS calibration for adverse pressure gradient flows.
\end{abstract}

\vspace*{0.5em}
\begin{IEEEkeywords}
Bayesian Optimization, RANS Calibration, GEKO Model, Turbulence Modelling.
\end{IEEEkeywords}

\section{INTRODUCTION}
Reynolds-averaged Navier-Stokes (RANS) models remain a cornerstone for industrial flow simulations due to their computational efficiency. These models are widely used in engineering applications, such as turbines, heat exchangers, and fluid transport systems, where accurate predictions of turbulent flows are critical. However, the standard practice of using default turbulence model coefficients often leads to suboptimal predictions, particularly in complex flows involving adverse pressure gradients (APG) and flow separation.

Predicting turbulent flow over a converging-diverging channel poses a significant challenge due to the complex interactions between turbulence and APG. While traditional RANS turbulence models, such as the $k$-$\omega$ Shear Stress Transport (SST) model \cite{menter1994} and the $k$-$\varepsilon$ model \cite{launder1974} offer computational efficiency, they frequently encounter difficulties in accurately predicting the locations of flow separation and reattachment. Large-eddy simulation and direct numerical simulation provide high-fidelity results but are computationally expensive, making them impractical for most industrial applications.

A key limitation of RANS models in APG flows is their inability to consistently capture flow separation and reattachment locations, leading to inaccuracies in the predicted skin friction coefficient. Studies have shown that two-equation eddy-viscosity models tend to either underpredict or overpredict flow separation, whereas Reynolds stress transport models generally improve predictions, particularly regarding the extent of separated flow regions \cite{Jesus2014LES, Jeyapaul2013}. Despite these improvements, RANS models struggle to fully capture the behavior of $C_f$ in APG regions with smooth bumps \cite{Menter2003}. For both attached and mildly separated flows, the skin friction coefficient exhibits similar behavior, indicating that APG effects, rather than separation alone, govern the primary flow physics \cite{Jesus2014LES, Jeyapaul2013}.

This study extends the turbo-RANS framework, originally introduced by McConkey et al. \cite{McConkey_Kalia_2024}, to improve the calibration of the Generalized $k$-$\omega$ (GEKO) model (as well as other turbulence closure models) for turbulent flow over a converging-diverging channel. The GEKO model provides additional flexibility by allowing the tuning of turbulence model coefficients to improve accuracy in adverse pressure gradient flows \cite{GEKO_model}. The calibration process follows a Bayesian optimization approach, systematically refining key turbulence model parameters ($C_\text{SEP}$ and $C_\text{NW}$) to minimize discrepancies between RANS predictions and high-fidelity reference data. Unlike previous studies that focused primarily on $C_p$ and $C_f$, this study also evaluates the predictive accuracy of RANS models on the streamwise velocity profiles at different Reynolds numbers.

The training data used for calibration consists of the coefficient of pressure and coefficient of friction from DNS at $Re = 12,600$. The calibrated model is then tested against streamwise velocity ($u$) profiles from both DNS ($Re = 12,600$) and LES ($Re = 20,580$), as well as the LES $C_f$ distribution. A key objective is to assess the robustness of the optimized coefficients across different Reynolds numbers, with the broader goal of enabling users to obtain a single set of optimal coefficients on one dataset and generalize them to different flow regimes. This approach enhances the practical applicability of turbo-RANS, allowing for turbulence model tuning that is transferable across moderate variations in Reynolds number and potentially minor geometric modifications. 

The remainder of this paper is structured as follows. Section~\ref{sec:turbulence_models} provides an overview of the turbulence models considered in this study, including the Generalized $k$-$\omega$ model. Section~\ref{sec:Methodology} outlines the computational setup and the turbo-RANS framework. Section~\ref{sec:Results} presents the results of the optimized turbulence model, including comparisons of $C_f$ and streamwise velocity ($u$) against DNS and LES benchmarks. Finally, Section~\ref{sec:Conclusion} summarizes the findings and discusses potential future applications.

\section{TURBULENCE MODELS IN RANS} \label{sec:turbulence_models}
Reynolds-averaged Navier-Stokes models solve for the Reynolds-averaged flow field, making them a viable option for computational applications in industry. However, their accuracy depends on the choice of turbulence closure model, which introduces empirical coefficients that influence the accuracy of the flow predictions. This section provides an overview of commonly used two-equation turbulence models, including the $k$-$\varepsilon$ and $k$-$\omega$ models, along with the more flexible Generalized $k$-$\omega$ model (GEKO).

The RANS equations are derived by decomposing the instantaneous velocity field into mean and fluctuating components. The resulting equations introduce the Reynolds stress term, which must be modeled to close the system. The general RANS equation is given as:
\begin{multline}
    \frac{\partial(\rho U_i)}{\partial t} 
    + \frac{\partial(\rho U_i U_j)}{\partial x_j} 
    = -\frac{\partial P}{\partial x_i} 
    + \frac{\partial}{\partial x_j} \bigg[ 
    \mu \left( \frac{\partial U_i}{\partial x_j} 
    + \frac{\partial U_j}{\partial x_i} \right) \\  
    - \rho \overline{u_i' u_j'} \bigg] \ ,
\end{multline}
\label{eq:RANS}
where $U_i$ is the $i$-th component of the time-averaged velocity, $\rho$ and $\mu$ are the density and dynamic viscosity of the fluid, $P$ is the pressure, and $\overline {u_i' u_j'}$ are components of the Reynolds stress tensor. The closure of the Reynolds stress term requires turbulence models, such as the $k$-$\varepsilon$, $k$-$\omega$, and GEKO models, which introduce additional transport equations for turbulence properties.

\subsection{Two-Equation Turbulence Models}
Two widely used RANS turbulence models are the \mbox{$k$-$\varepsilon$} and $k$-$\omega$ models. The $k$-$\varepsilon$ model \cite{launder1974} is known for its robustness and efficiency in industrial applications. It solves transport equations for the turbulent kinetic energy ($k$) and the turbulence dissipation rate ($\varepsilon$), with turbulent viscosity defined based on these quantities. While effective in free-stream flows, it struggles with accuracy in near-wall regions, particularly in adverse pressure gradient (APG) flows.

The $k$-$\omega$ model \cite{wilcox1988} instead introduces the specific dissipation rate ($\omega$) to govern the scale of turbulent eddies. This model provides improved accuracy in near-wall regions but can be sensitive to free-stream turbulence levels. Both models serve as foundational RANS closures, with modifications such as the GEKO model introduced to enhance predictive accuracy in complex flow conditions.

\subsection{GEKO Turbulence Model}

The Generalized $k$-$\omega$ model is an extension of the standard $k$-$\omega$ model, designed to provide greater flexibility in turbulence modeling by incorporating additional tunable coefficients \cite{GEKO_model}. Unlike conventional turbulence models that rely on fixed empirical constants, the GEKO model introduces parameters that allow users to modify the model's behavior based on specific flow conditions. This adaptability makes it particularly suitable for flows involving adverse pressure gradients, separation, and reattachment, such as the converging-diverging channel examined in this study.

The governing equations of the GEKO model are built upon the standard $k$-$\omega$ formulation, with additional blending functions that allow for user-defined modifications. The transport equations for turbulent kinetic energy ($k$) and the specific dissipation rate ($\omega$) are given as
\begin{align}
\frac{\partial(\rho k)}{\partial t} 
+ \frac{\partial(\rho U_j k)}{\partial x_j} 
&= P_k - C_\mu \rho k \omega \notag \\
&\quad + \frac{\partial}{\partial x_j} \left[ \left( \mu + \frac{\mu_t}{\sigma_k} \right) 
\frac{\partial k}{\partial x_j} \right] \ ,
\label{eq:geko_k_transport}
\end{align}
\begin{multline}
    \frac{\partial(\rho \omega)}{\partial t} 
    + \frac{\partial(\rho U_j \omega)}{\partial x_j} 
    = C_{\omega 1} F_1 \frac{\omega}{k} P_k 
    - C_{\omega 2} F_2 \rho \omega^2 
    + \rho F_3 C_D \\
    + \frac{\partial}{\partial x_j} 
    \left[ \left( \mu + \frac{\mu_t}{\sigma_\omega} \right) 
    \frac{\partial \omega}{\partial x_j} \right] \ ,
\label{eq:geko_omega_transport}
\end{multline}
where $F_1$, $F_2$, and $F_3$ are blending functions that allow for user-defined modifications to the model, affecting different regions of the flow.

\subsubsection{Tunable Coefficients in the GEKO Model}

The GEKO model incorporates six tunable coefficients, allowing users to modify specific aspects of turbulence behavior to better match different flow conditions \cite{GEKO_model}. These coefficients influence boundary-layer separation, near-wall turbulence, shear-flow mixing, jet spreading, secondary flows, and curvature effects. The key tunable parameters are:

\begin{itemize} 
\item $C_\text{SEP}$: This coefficient primarily controls boundary-layer separation by regulating eddy viscosity in regions with APG. A higher value reduces eddy viscosity, making the flow more susceptible to separation, while a lower value increases viscosity, encouraging flow attachment. 
\item $C_\text{NW}$: This coefficient adjusts near-wall turbulence characteristics, influencing wall-shear stress and heat transfer. Increasing $C_\text{NW}$ enhances shear stress at the wall, improving accuracy in non-equilibrium boundary layers. 
\item $C_\text{MIX}$: This coefficient governs turbulent mixing in shear layers, while $C_\text{JET}$ modifies jet-spreading rates. $C_\text{CORNER}$ accounts for secondary-flow effects, and $C_\text{CURV}$ enhances accuracy in cases where strong streamline curvature influences turbulence. These parameters provide additional flexibility in adjusting model behavior for specific flow conditions. 
\end{itemize}

\subsubsection{Application of GEKO in this Study}

In this study, the turbo-RANS framework is employed to optimize the values of $C_\text{SEP}$ and $C_\text{NW}$, with the objective of enhancing the accuracy of boundary-layer separation, wall-shear stress, and pressure recovery predictions. The optimization process is guided by pressure recovery data from DNS and LES benchmarks, ensuring that the calibrated GEKO model provides more reliable results across different Reynolds numbers.

A Python-based automation script was used to modify the Ansys Fluent journal file for each optimization iteration. The first set of iterations was conducted using a Sobol sequence to explore the parameter space, followed by Bayesian optimization to refine the coefficients further. The upper and lower bounds of $C_\text{SEP}$ and $C_\text{NW}$ were set according to the recommended GEKO model limits:
\begin{equation}
0.70 \leq C_\text{SEP} \leq 2.5 \ , \quad -2 \leq C_\text{NW} \leq 2 \ .
\label{eq:geko_bounds}
\end{equation}

The calibrated coefficients were then tested at higher Reynolds numbers to assess their generalizability. Moreover, using the coefficient of pressure ($C_p$) and coefficient of friction ($C_f$) as training data, the model's performance at $Re = 12,600$ was further evaluated by comparing the predicted streamwise velocity to DNS results. The findings demonstrate that tuning $C_\text{SEP}$ and $C_\text{NW}$ significantly enhances RANS predictions in adverse pressure gradient flows, particularly improving the accuracy of streamwise velocity and $C_f$ trends. The final optimized coefficients are compared to baseline results in Section~\ref{sec:Results}.

The flexibility of the GEKO model, combined with Bayesian optimization, allows for a more systematic and automated approach to turbulence model calibration. This study reinforces the effectiveness of data-driven turbulence model tuning for improving predictive accuracy in complex flow configurations.

\section{METHODOLOGY}\label{sec:Methodology}

\subsection{Computational Setup}

\begin{figure}[htbp]
    \centering
    \includegraphics[width=0.9\linewidth]{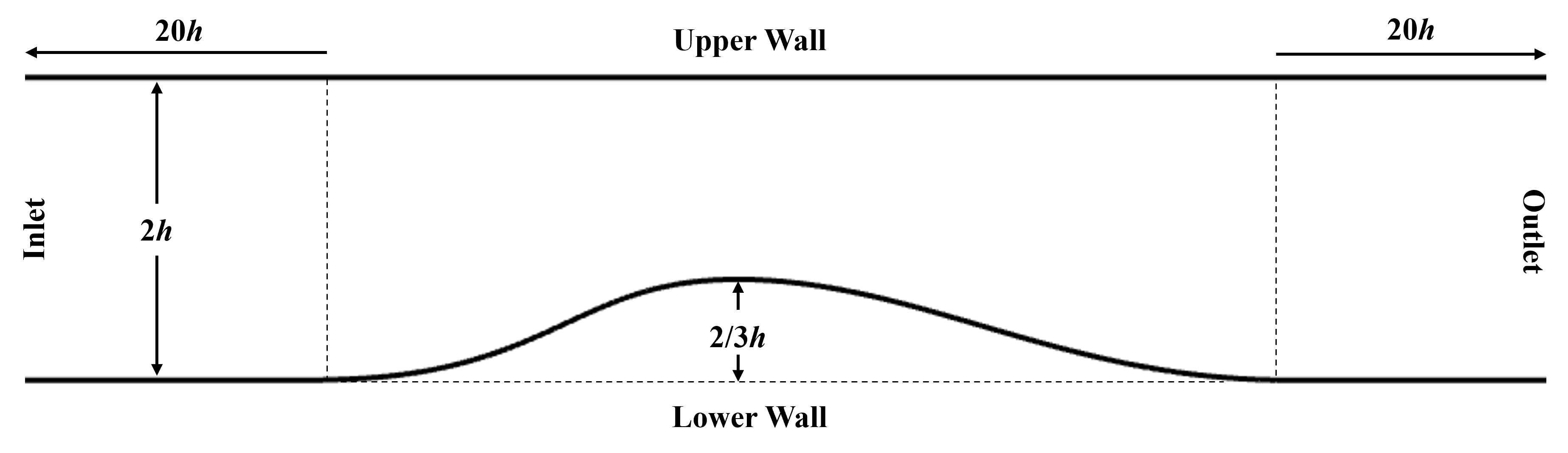}
    \caption{Computational domain for the converging-diverging channel.}
    \label{fig:domain_channel}
\end{figure}

\begin{figure}[htbp]
    \centering
    \includegraphics[width=0.75\linewidth]{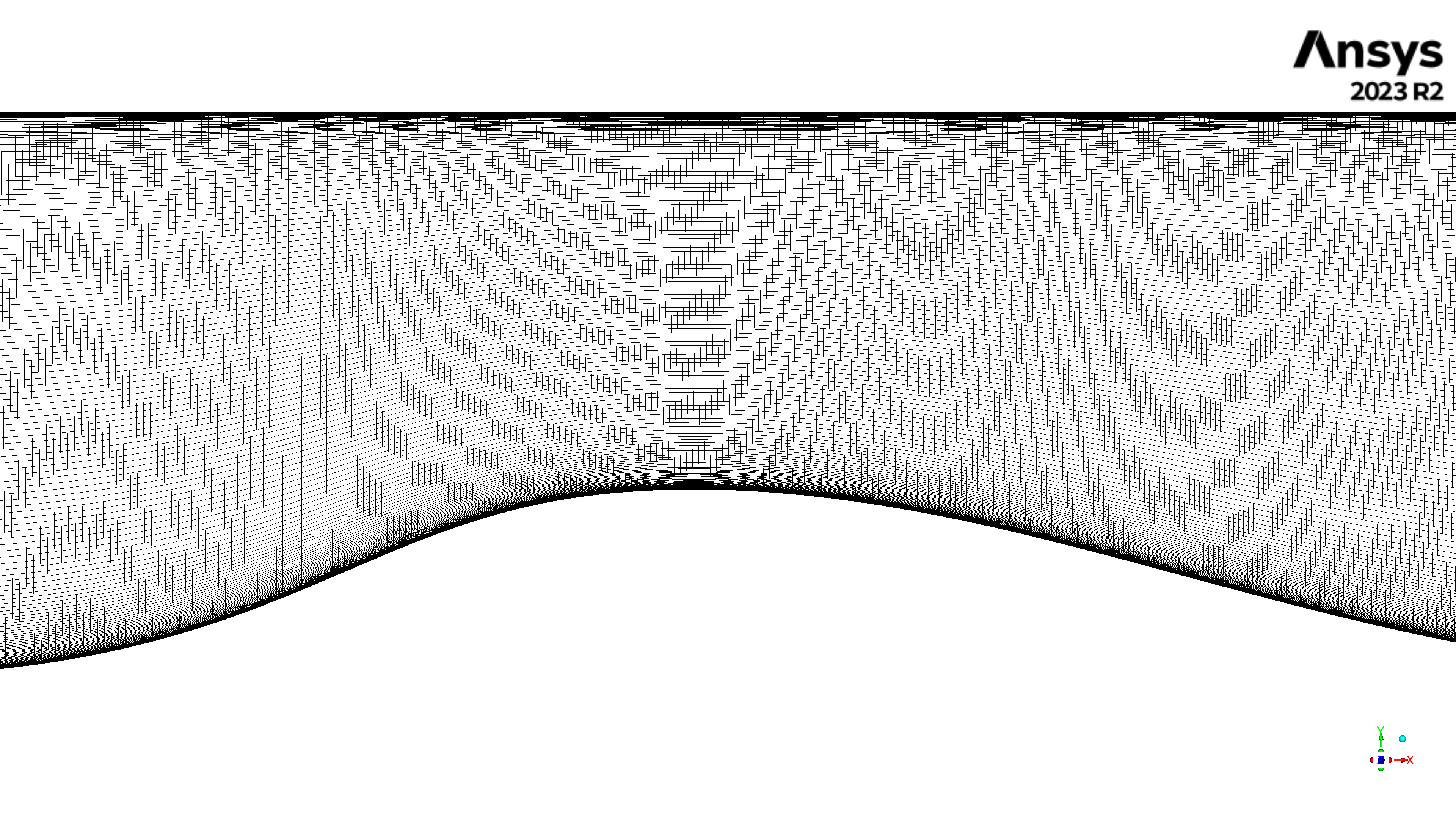}
    \caption{Structured mesh used for the RANS simulations of the converging-diverging channel.}
    \label{fig:mesh_cdc}
\end{figure}

The numerical simulations are conducted for steady-state, incompressible, turbulent flow through a converging-diverging channel using the GEKO turbulence model within the Reynolds-averaged Navier-Stokes framework. The baseline Reynolds number. defined using the maximum streamwise velocity ($U_\text{max}$) and the half-channel height ($h$), is $Re = 12,600$. The computational domain extends $20h$ upstream and downstream of the bump to ensure fully-developed inlet flow conditions and accurate outflow predictions. The domain geometry and structured mesh used in the simulations are shown in Figs.~\ref{fig:domain_channel} and~\ref{fig:mesh_cdc}, respectively.

The DNS reference data for this case is obtained from Marquillie et al.~\cite{Marquillie2008,Marquillie2011}, while the structured computational mesh is adapted from McConkey et al.~\cite{McConkey_Kalia_2024}. The mesh consists of 610,720 grid cells, which has been validated for mesh independence in the converging-diverging region~\cite{McConkey_Kalia_2024}.

\subsection{Reynolds-number Variation}

To assess the influence of Reynolds number on the flow predictions, additional simulations are performed at $Re = 20,580$. This is achieved by modifying the kinematic viscosity while maintaining the same characteristic length ($h$) and velocity scale ($U_\text{max}$). Towards this objective, the updated kinematic viscosity, $\nu_2$, for the higher Reynolds number is computed using
\begin{equation}
\nu_2 = \nu_1 \cdot \frac{Re_1}{Re_2} \ ,
\label{eq:final_re}
\end{equation}
where the baseline viscosity is $\nu_1 = 6.7063 \times 10^{-5} \ \text{m}^2 \text{ s}^{-1}$ at $Re_1 = 12,600$, and for $Re_2 = 20,580$, the updated viscosity is $\nu_2 = 4.1059 \times 10^{-5} \ \text{m}^2 \text{ s}^{-1}$. This approach ensures a controlled Reynolds number variation without altering the geometry or inlet velocity.

\subsection{Boundary Conditions and Solver Settings}

Boundary conditions and fluid properties are selected to match the DNS reference simulation. 
The inlet plane is positioned \(20h\) upstream of the bump, where a Neumann pressure boundary condition (zero normal pressure gradient) is applied. The inflow velocity profile is uniform with \( U_\text{max} = 0.845 \) m s\(^{-1}\), ensuring consistency with the mass flow rate used in previous RANS studies~\cite{McConkey_Kalia_2024}. The turbulence variables at the inlet are prescribed as 
\( k = 4.28421 \times 10^{-4} \) m\(^2\) s\(^{-2}\) and \( \omega = 0.26993 \) s\(^{-1}\).

At the outlet, a fixed gauge pressure of zero is applied, while a zero-gradient boundary condition is imposed for all other quantities. The outlet plane is positioned \( 20h \) downstream of the bump to avoid numerical artifacts. A no-slip boundary condition is enforced on all walls.

For the baseline simulation, the fluid is assumed to have a density of \( \rho = 1 \) kg m$^{-3}$ and a kinematic viscosity of \( \nu = 6.7063 \times 10^{-5} \) m$^2$ s$^{-1}$, corresponding to a Reynolds number of \( Re = 12,600 \). The simulations are carried out using the Semi-Implicit Method for Pressure Linked Equations-Consistent (SIMPLEC) algorithm for pressure-velocity coupling. The convective terms in the momentum equation are discretized using a second-order upwind scheme, while a first-order upwind scheme is applied to the turbulence transport equations. Diffusion terms are discretized with a second-order central-difference scheme.

\subsection{Wall Treatment and Grid Resolution}

\begin{figure}[htbp]
    \centering
    \includegraphics[width=0.48\textwidth]{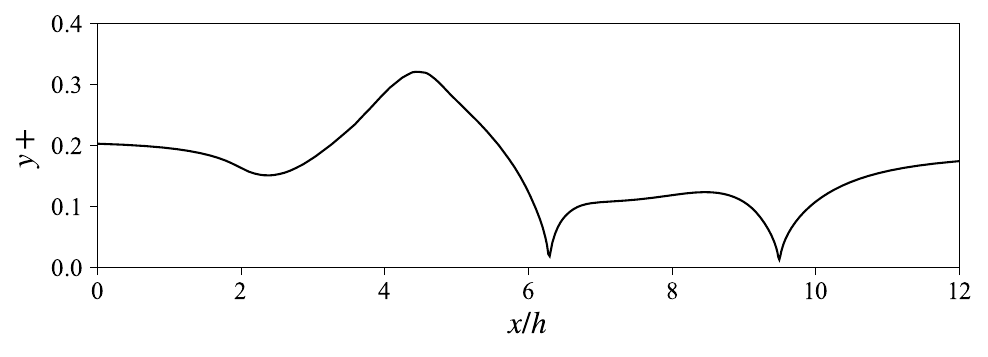}
    \caption{Wall treatment verification, ensuring $y^+ < 1$.}
    \label{fig:y+}
\end{figure}

A fine grid is used to ensure accurate modeling of near-wall turbulence. The grid is refined such that the first grid point lies within the viscous sublayer, maintaining a non-dimensional wall distance $y^+ < 1$ throughout the domain. The $y^+$ value is defined as
\begin{equation}
y^+ = \frac{u_\tau y}{\nu},
\label{eq:y+_eq}
\end{equation}
where $u_\tau$ is the friction velocity at the wall, $y$ is the normal distance from the wall to the first cell center, and $\nu$ is the kinematic viscosity. The distribution of $y^+$ values across the computational domain is shown in Fig.~\ref{fig:y+}, confirming that the boundary layer is well-resolved without requiring wall functions.

\subsection{High-fidelity Data for Comparison}

To further assess the accuracy of the optimized GEKO model, both direct numerical simulation (DNS) and large-eddy simulation (LES) datasets are used as benchmark data. The DNS dataset at $Re = 12,600$ \cite{Marquillie2008, Marquillie2011} is employed as training data for model calibration, where the coefficient of pressure and coefficient of friction serve as the reference quantities. After calibration, the optimized turbulence model is tested against additional flow characteristics to evaluate its predictive performance. 

At $Re = 12,600$, the streamwise velocity profiles from DNS are used to verify whether the model retains accuracy in velocity field predictions. Further, to assess the robustness of the calibrated model beyond its training dataset, its predictions are evaluated at a higher Reynolds number of $Re = 20,580$ against LES results from Schiavo et al.~\cite{LES_data}. In this case, both the streamwise velocity and coefficient of friction are compared to determine the model's ability to generalize across different flow conditions.

The reference data for both DNS and LES were sourced from McConkey et al.~\cite{McConkeySciDataPaper2021}, which compiled high-fidelity datasets by combining the DNS data from Laval~\cite{NASA_DNS_Data_2024} and the LES data from Schiavo et al.~\cite{LES_data}. This dataset provides a consolidated benchmark for evaluating RANS models, ensuring consistency in comparison across different Reynolds numbers.

By using LES data for validation at a higher Reynolds number, this study ensures that the optimized turbulence model maintains accuracy beyond the conditions it was trained on. This enables an assessment of the generalizability of the calibrated GEKO model coefficients, determining whether a single optimized set of parameters can provide reliable predictions across similar flow regimes. The results of this evaluation are presented in Section~\ref{sec:Results}.

\section{RESULTS AND DISCUSSION} \label{sec:Results}

\subsection{Automation of Coefficient Calibration}

The turbo-RANS framework was used to optimize the coefficients of the GEKO turbulence model for the converging-diverging channel case. A Python script was developed to automate the workflow by executing an Ansys Workbench journal file, which set up the Fluent simulation from mesh import to the export of RANS predictions. The script modified the journal file to iteratively update the values of $C_\text{SEP}$ and $C_\text{NW}$ based on the optimization process prescribed by the turbo-RANS algorithm.

The calibration process was driven by the Generalized Error and Default Coefficient Preference (GEDCP) objective function, which was formulated using the coefficient of pressure and coefficient of friction from the DNS dataset at $Re = 12,600$. The first iteration used the default GEKO model coefficients, followed by ten Sobol-sampled iterations to explore the parameter space. Subsequently, 20 Bayesian optimization iterations were conducted to refine the coefficient values. After 30 iterations, the optimal values for $C_\text{SEP}$ and $C_\text{NW}$ were identified as those that minimized the GEDCP objective function, ensuring improved alignment of RANS predictions with high-fidelity reference data.

\begin{figure*}[htbp]
    \centering
    \subfloat[DNS]{%
        \includegraphics[width=0.32\textwidth]{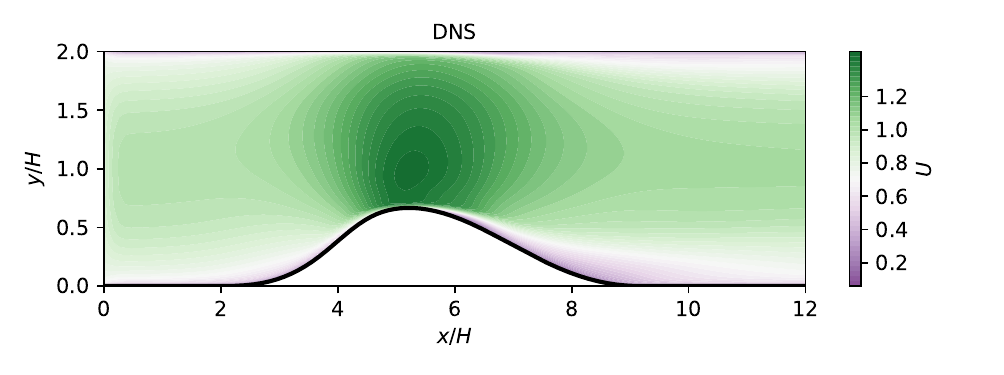}
        \label{fig:dns}
    }%
    \hfill
    \subfloat[GEKO]{%
        \includegraphics[width=0.32\textwidth]{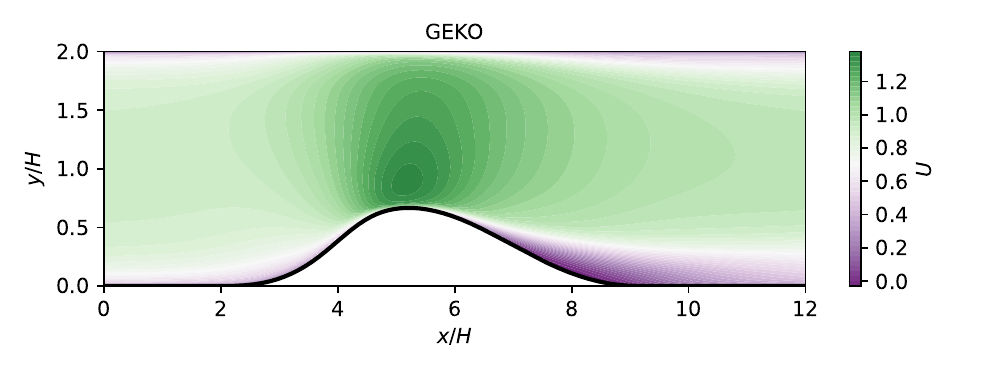}
        \label{fig:geko}
    }%
    \hfill
    \subfloat[GEKO (turbo-RANS)]{%
        \includegraphics[width=0.32\textwidth]{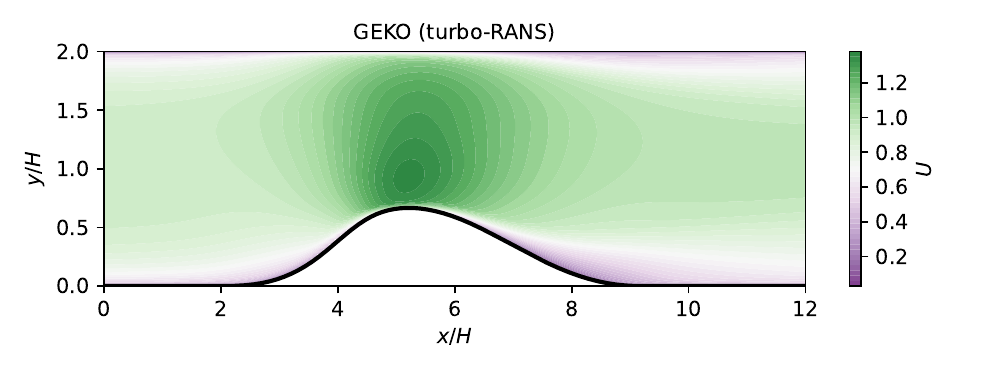}
        \label{fig:geko_turbo}
    }%
    \caption{Contours of the streamwise mean velocity of $U$ in a vertical plane of the converging-diverging channel.}
    \label{fig:velocity_comparison}
\end{figure*}

\begin{figure*}[htbp]  
    \centering
    \subfloat[Comparison of mean streamwise velocity vertical profiles at different $x$-locations in a converging-diverging channel at $Re = 12,600$.]{%
        \includegraphics[width=0.7\textwidth]{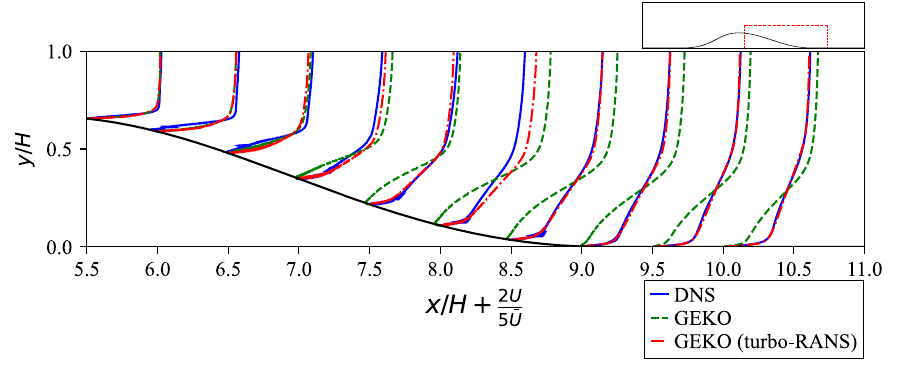}
        \label{fig:velocity_profiles_dns}
    }%
    \hfill
    \subfloat[Comparison of mean streamwise velocity vertical profiles at different $x$-locations in a converging-diverging channel at $Re = 20,580$.]{%
        \includegraphics[width=0.7\textwidth]{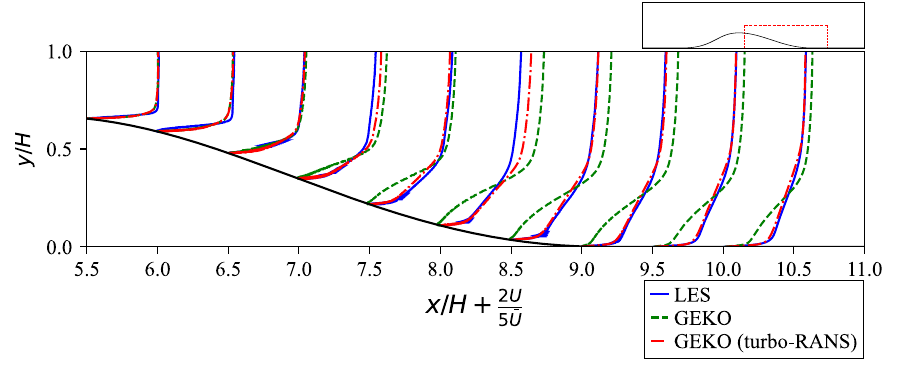}
        \label{fig:velocity_profiles_les}
    }%
    \caption{Comparison of velocity vertical profiles at various streamwise locations.}
    \label{fig:velocity_profiles_comparison}
\end{figure*}

\subsection{Optimization Problem}

For the converging-diverging channel case, the same optimization framework as described by McConkey et al.~\cite{McConkey_Kalia_2024} was utilized. The (GEDCP) objective function was used to calibrate the GEKO model coefficients, $C_\text{SEP}$ and $C_\text{NW}$, based on sparse wall-pressure measurements. The GEDCP function optimizes these coefficients by minimizing a regularized version of the mean absolute percentage error (MAPE) between the computed and reference values of the pressure coefficient.

The sparse reference dataset consists of 10 measurement points along the bottom wall of the channel, selected to capture key flow regions, including the converging and diverging sections of the channel. A complete description of the GEDCP function, including its mathematical formulation and implementation, is provided in McConkey et al.~\cite{McConkey_Kalia_2024}. The weighting factor $\lambda_{\text{coef}} = 0.25$ was applied to penalize deviations from the default GEKO coefficients.

The Bayesian optimization process follows the same methodology outlined in the original paper. It begins with an initial Sobol sequence sampling to explore the parameter space, followed by iterative refinement using a Gaussian process surrogate model. The optimization process efficiently balances exploration and exploitation to converge on the optimal values of $C_\text{SEP}$ and $C_\text{NW}$. The coefficient bounds used for these two coefficients in the optimization were initially set according to the Ansys GEKO recommendations~\cite{GEKO_model}.

However, during the Bayesian optimization process, the optimization consistently converged towards the lower bound of \(C_{\text{SEP}} = 0.7\). Given that there was no specific justification for restricting the lower bound for this parameter to 0.7 in the Ansys documentation, we further reduced the lower bound to 0.3, allowing for more exploration of lower eddy-viscosity behaviors in adverse pressure-gradient flows. This adjustment ultimately led to improved performance in matching the reference dataset.

Since the final objective of this study was to derive a single generalizable set of optimized coefficients applicable to different Reynolds numbers, the final values were obtained by optimizing for both \(C_p\) and \(C_f\) at \(Re = 12,600\). The resulting optimal coefficients were
\begin{equation}
C_\text{SEP} = 0.489 \ , \quad C_\text{NW} = 1.778 \ .
\end{equation}
These values were then used for subsequent comparison of the streamwise velocity at \(Re = 12,600\) and \(Re = 20,580\) between the DNS/LES reference data and the RANS model predictions in order to assess their robustness across different flow conditions.

\subsection{Improved Reattachment Prediction with turbo-RANS}
Figure~\ref{fig:velocity_comparison} displays the scalar field of the streamwise mean velocity $U$ in a vertical plane through the converging-diverging channel for the reference DNS data, as well as for the RANS model predictions of this quantity obtained using the standard GEKO model and the recalibrated GEKO (turbo-RANS) model. A careful inspection shows qualitatively that the streamwise mean velocity predictions obtained using GEKO (turbo-RANS) are in better conformance with the DNS results than those obtained using GEKO.

Figure~\ref{fig:velocity_profiles_comparison} presents a comparison of the vertical profiles of the streamwise velocity at multiple streamwise locations $x/h$ for both the DNS dataset at $Re = 12,600$ (Figure~\ref{fig:velocity_profiles_dns}) and the LES dataset at $Re = 20,580$ (Figure~\ref{fig:velocity_profiles_les}). The baseline GEKO turbulence closure model is prone to overpredicting the level of separation and faces challenges in accurately determining reattachment, particularly in the region of the flow experiencing an adverse pressure gradient downstream of the bump. In contrast, the GEKO (turbo-RANS) model demonstrates improved accuracy in predicting the strreamwise mean velocity profiles, particularly in the reattachment region. To improve visualization and facilitate a direct comparison between different datasets, the velocity profiles are scaled using a modified coordinate transformation based on the local mean velocity.

The optimized coefficients ($C_{\text{SEP}}$ and $C_{\text{NW}}$) at $Re = 12,600$, successfully generalize to the higher Reynolds number test case ($Re = 20,580$), as observed in Figure~\ref{fig:velocity_profiles_les}. This demonstrates the transferability of the optimized coefficients, making turbo-RANS a practical approach for improving RANS predictions beyond the calibration dataset. The improved reattachment location and velocity distribution suggest that the optimized model better captures flow recovery after separation, addressing a critical limitation of conventional RANS models.

\begin{figure}[htbp]
    \centering
    \includegraphics[width=0.48\textwidth]{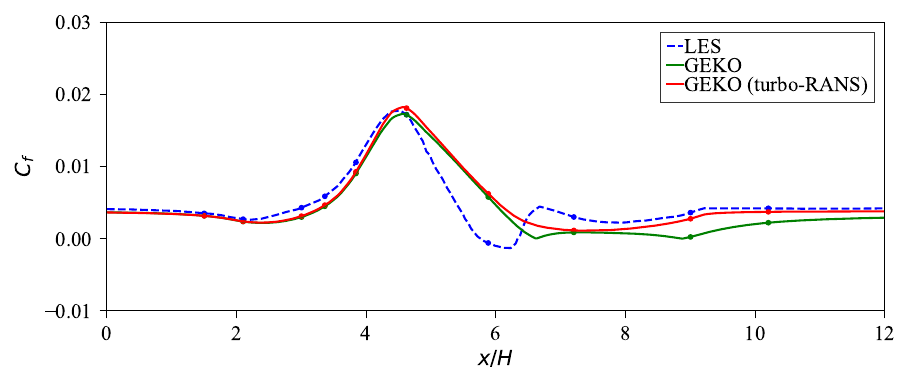}
    \caption{Comparison of skin friction coefficient at $Re = 20,580$.}
    \label{fig:cf_comparison}
\end{figure}

Figure~\ref{fig:cf_comparison} compares the $C_f$ predictions obtained using the GEKO and GEKO (turbo-RANS) models against reference LES data. While the GEKO (turbo-RANS) model provides some improvement over the baseline GEKO model, the gain is not as evident as in the streamwise velocity profiles. This is largely due to the inherent limitations of two-equation turbulence models, which have difficulty in accurately resolving wall-shear stress, particularly in adverse pressure gradient flows. The standard RANS eddy-viscosity assumption tends to overpredict or underpredict separation and reattachment trends, leading to discrepancies in $C_f$ distributions. Although the optimized coefficients in GEKO (turbo-RANS) model improve the predictive capability to some extent, fundamental limitations of the turbulence closure formulation restrict further accuracy gains. Despite these challenges, the slight improvement in $C_f$ suggests that Bayesian optimization is, nevertheless, useful in refining turbulence model performance for specific cases.

\section{Conclusion}\label{sec:Conclusion}  

This study applied the turbo-RANS framework to optimize the GEKO turbulence model for a converging-diverging channel, using Bayesian optimization with DNS data at \(Re = 12,600\). The optimized model showed significant improvements in the streamwise mean velocity predictions, particularly in capturing reattachment, a key limitation of standard (conventional) RANS models. The calibrated coefficients generalize well across different Reynolds numbers for the converging-diverging channel flow, demonstrating robustness in different flow conditions.

However, improvements in \(C_f\) predictions remain limited due to the challenges of two-equation turbulence models in accurately resolving wall-shear stress. While Bayesian optimization refines turbulence model behavior, inherent RANS closure limitations restrict further accuracy gains. Nonetheless, turbo-RANS provides a systematic, data-driven approach to turbulence model calibration without modifying the underlying closure formulation. Future work will explore extensions to other adverse pressure gradient flows using additional reference datasets.

\bibliographystyle{IEEEtran}
\bibliography{references}

\begin{thebibliography}{10}
\providecommand{\url}[1]{#1}
\csname url@samestyle\endcsname
\providecommand{\newblock}{\relax}
\providecommand{\bibinfo}[2]{#2}
\providecommand{\BIBentrySTDinterwordspacing}{\spaceskip=0pt\relax}
\providecommand{\BIBentryALTinterwordstretchfactor}{4}
\providecommand{\BIBentryALTinterwordspacing}{\spaceskip=\fontdimen2\font plus
\BIBentryALTinterwordstretchfactor\fontdimen3\font minus \fontdimen4\font\relax}
\providecommand{\BIBforeignlanguage}[2]{{%
\expandafter\ifx\csname l@#1\endcsname\relax
\typeout{** WARNING: IEEEtran.bst: No hyphenation pattern has been}%
\typeout{** loaded for the language `#1'. Using the pattern for}%
\typeout{** the default language instead.}%
\else
\language=\csname l@#1\endcsname
\fi
#2}}
\providecommand{\BIBdecl}{\relax}
\BIBdecl

\bibitem{menter1994}
F.~R. Menter, ``Two-equation eddy-viscosity turbulence models for engineering applications,'' \emph{AIAA Journal}, vol.~32, no.~8, pp. 1598--1605, 1994.

\bibitem{launder1974}
B.~E. Launder and D.~B. Spalding, ``The numerical computation of turbulent flows,'' \emph{Computer Methods in Applied Mechanics and Engineering}, vol.~3, no.~2, pp. 269--289, 1974.

\bibitem{Jesus2014LES}
A.~B. Jesus, L.~A. C.~A. Schiavo, J.~L.~F. Azevedo, and W.~R. Wolf, ``Large eddy simulations of turbulent flows with adverse pressure gradients,'' in \emph{29th Congress of the International Council of the Aeronautical Sciences (ICAS)}, St. Petersburg, Russia, September 7--12 2014.

\bibitem{Jeyapaul2013}
E.~Jeyapaul and C.~Rumsey, ``Analysis of highly-resolved simulations of 2-d humps toward improvement of second-moment closures.''\hskip 1em plus 0.5em minus 0.4em\relax Grapevine, TX: AIAA, January 2013.

\bibitem{Menter2003}
F.~R. Menter, M.~Kuntz, and R.~Langtry, ``{Ten years of industrial experience with the SST turbulence model},'' \emph{Turbulence, Heat and Mass Transfer 4}, pp. 625--632, 2003.

\bibitem{McConkey_Kalia_2024}
R.~McConkey, N.~Kalia, E.~Yee, and F.-S. Lien, ``{Turbo-RANS: Straightforward and efficient Bayesian optimization of turbulence model coefficients},'' \emph{International Journal of Numerical Methods for Heat and Fluid Flow}, 2024.

\bibitem{GEKO_model}
F.~R. Menter and R.~Lechner, ``{Best practice: Generalized k-{$\omega$} (GEKO) two-equation turbulence modeling in Ansys CFD},'' ANSYS, Tech. Rep., 2021.

\bibitem{wilcox1988}
D.~C. Wilcox, ``{Reassessment of the scale-determining equation for advanced turbulence models},'' \emph{AIAA Journal}, vol.~26, no.~11, pp. 1299--1310, 1988.

\bibitem{Marquillie2008}
M.~Marquillie, J.~P. Laval, and R.~Dolganov, ``{Direct numerical simulation of a separated channel flow with a smooth profile},'' \emph{Journal of Turbulence}, vol.~9, pp. 1--23, 2008.

\bibitem{Marquillie2011}
M.~Marquillie, U.~Ehrenstein, and J.~P. Laval, ``{Instability of streaks in wall turbulence with adverse pressure gradient},'' \emph{Journal of Fluid Mechanics}, vol. 681, pp. 205--240, 2011.

\bibitem{LES_data}
L.~A. Schiavo, A.~B. Jesus, J.~L. Azevedo, and W.~R. Wolf, ``{Large eddy simulations of convergent-divergent channel flows at moderate Reynolds numbers},'' \emph{International Journal of Heat and Fluid Flow}, vol.~56, pp. 137--151, 12 2015.

\bibitem{McConkeySciDataPaper2021}
\BIBentryALTinterwordspacing
R.~McConkey, E.~Yee, and F.~S. Lien, ``{A curated dataset for data-driven turbulence modelling},'' \emph{Scientific Data}, vol.~8, no.~1, pp. 1--14, 2021. [Online]. Available: \url{http://dx.doi.org/10.1038/s41597-021-01034-2}
\BIBentrySTDinterwordspacing

\bibitem{NASA_DNS_Data_2024}
J.~Laval. (2024) Dns: 2-d converging-diverging channel, re=12600. Available at: \href{https://turbmodels.larc.nasa.gov/Other_DNS_Data/conv-div-channel12600.html}{turbmodels.larc.nasa.gov/...}

\end{thebibliography}

\end{document}